\documentclass[aps,prd,longbibliography,twocolumn,nofootinbib,10pt]{revtex4-2}
\date{\today}

\usepackage{amsmath}
\usepackage{amsfonts}
\usepackage{graphicx}
\usepackage{dsfont}
\usepackage[german, english]{babel}
\usepackage{amssymb}
\usepackage{ifthen}
\usepackage{soul}
\usepackage{color}
\usepackage{xcolor}

\usepackage{bm}

\newcommand{\secref}[1]{Section \ref{#1}}
\newcommand{\figref}[1]{Fig. \ref{#1}}
\renewcommand{\eqref}[1]{Eq. (\ref{#1})}
\newcommand{\ii}{\mathrm{i}}

\RequirePackage[colorlinks=true,allcolors=blue,
unicode=true,hypertexnames=false]{hyperref}
\hypersetup{pdfstartview={XYZ null null 1.25}}

\graphicspath{{images/}}

\begin{document}

\title{
Microscopic derivation of superconductor-insulator boundary conditions for Ginzburg-Landau theory revisited. Enhanced superconductivity at boundaries with and without magnetic field
}

\author{Albert Samoilenka}
\affiliation{Department of Physics, KTH-Royal Institute of Technology, SE-10691, Stockholm, Sweden}

\author{Egor~Babaev}
\affiliation{Department of Physics, KTH-Royal Institute of Technology, SE-10691, Stockholm, Sweden}

\begin{abstract}
Using the standard Bardeen-Cooper-Schrieffer (BCS) theory, we revise microscopic derivation of the superconductor-insulator boundary conditions for the Ginzburg-Landau (GL) model. We obtain a negative contribution to free energy in the form of surface integral. Boundary conditions for the conventional superconductor have the form $\textbf{n} \cdot \nabla \psi = \text{const} \psi$. These are shown to follow from considering the order parameter reflected in the boundary. The boundary conditions are also derived for more general GL models with higher-order derivatives and pair-density-wave states. It shows that the boundary states with higher critical temperature and the boundary gap enhancement, found recently in BCS theory, are also present in microscopically-derived GL theory. In the case of an applied external field, we show that the third critical magnetic-field value $H_{c3}$ is higher than what follows from the de Gennes boundary conditions and is also significant in type-I regime.
\end{abstract}

\maketitle

\section{Introduction}
Superconductivity in the Ginzburg-Landau (GL) model \cite{Ginzburg1950} is described by a complex-valued field $\psi(\textbf{r})$, which is called an order parameter or gap.
In the bulk of the sample, $\psi$ is found as a minimum of the free-energy functional $F_{bulk}[\psi]$.
The form of this functional was microscopically derived first by Gor'kov \cite{Gorkov1959}.
To solve for $\psi$ near a boundary of superconductor one has to take into account the influence of the material outside the sample.
This is done by a microscopically derived boundary condition for $\psi$ or, equivalently, by an additional surface term $F_{surf}[\psi]$ in free-energy functional.
Namely, dropping vector potential (it is restored in \secref{sec_GL_H}), the free energy $F$ of a superconductor placed in $\Omega$ is given by:
\begin{equation}\label{F_GL}
\begin{gathered}
F = F_{bulk} + F_{surf} \\
F_{bulk} = \int_{\Omega} d\textbf{r} \left[ K |\nabla \psi|^2 + \alpha |\psi|^2 + \beta |\psi|^4 \right] \\
F_{surf} = \gamma \int_{\partial \Omega} d\textbf{r}_{||} |\psi|^2
\end{gathered}
\end{equation}

The order parameter is found as a minimum of $F$.
Hence the boundary condition at $\partial \Omega$ is:
\begin{equation}\label{bc_Lambda}
\textbf{n} \cdot \nabla \psi = - \frac{\gamma}{K} \psi
\end{equation}
where $\textbf{n}$ is unit vector pointing outside of the sample.

We begin by reviewing how the value of $\gamma$ changes the gap behavior near the boundary in different systems.
A calculation by de Gennes \cite{deGennes_Boundary,book_de_Gennes} gave that $\gamma > 0$ at a boundary between a superconductor and a normal metal.
Hence superconductivity is suppressed near such a boundary.
One also defines the extension length $\Lambda = K / \gamma > 0$, which controls the range of induced superconducting correlation in the metal.
On the other hand when $\gamma < 0$ superconductivity is enhanced.
Such a situation is realized in several cases. Namely, for a contact with a superconductor with a higher critical temperature \cite{fink1969surface}, for a local increase of electron-phonon coupling constants near the surface \cite{ginzburg1964surface} and for superconductivity on twinning planes \cite{buzdin1981localized,khlyustikov1987twinning}.
A phenomenological model with $\gamma < 0$ was analyzed in a number of works, see e.g. \cite{fink1969surface,buzdin1981localized,andryushin1993boundary,ivanov1987heat,abrikosov1989thermodynamic,averin1983theory,simonin1986surface,kaganov1972phenomenological}.
Boundary conditions for the interface between superconductor and insulator were studied microscopically in \cite{abrikosov1965concerning,deGennes_Boundary,CdGM_french,book_de_Gennes}, yielding the conclusion that it is a good approximation to set $\gamma = 0$.
We will call the corresponding boundary conditions the de Gennes boundary conditions.

However, the situation for the superconductor-insulator interface is not trivial.
Namely, it was recently shown microscopically that boundaries of superconductors can have (i) higher critical temperature and (ii) the gap can be enhanced at the scale of the bulk coherence length.
We call this enhancement of superconductivity \footnote{Note, that, near a superconductor-insulator boundary, the gap oscillates with period $\simeq 1 / k_F$, where $k_F$ is the Fermi momentum.
Some works term the first peak of these Friedel oscillations "enhancement" \cite{giamarchi1990onset,stojkovic1993order}, even in the case where the averaged gap is not enhanced, which corresponds to the situation where de Gennes boundary conditions apply in a GL theory.
Here we mean by enhancement an increase in the gap averaged over a length scale that is much larger than $1 / k_F$.} the \textit{boundary state}.
It was found in one, two, and three-dimensional superconductors in the tight-binding BCS model and for the one-dimensional continuous BCS model \cite{samoilenka2020boundary}.
An earlier study of the standard three-dimensional continuous BCS model concluded that the boundary state is absent: the averaged gap near the boundaries is neither suppressed nor enhanced \cite{giamarchi1990onset} or weakly suppressed \cite{stojkovic1993order}.
Coulomb repulsion-induced boundary states were reported to form in certain cases in a three-dimensional continuous model where the interaction was attractive for small energies and repulsive for higher energies \cite{giamarchi1990onset}.
Superconductivity enhancement in the form of pair-density-wave (PDW) boundary states was found in spin-imbalanced superconductors \cite{barkman2019surface,samoilenka2020pair} as well.

The situation is also controversial experimentally.
Evidence for a substantially enhanced superconductivity near the boundary was reported in some elemental and high-temperature superconductors, see, e.g. \cite{fink1969surface,lortz2006origin, janod1993split, butera1988high,tsindlekht2004tunneling,belogolovskii2010zirconium,khasanov2005anomalous,khlyustikov2011critical,khlyustikov2016surface,mangel2020stiffnessometer}.
The enhanced surface superconductivity was also mentioned in the context of enhancement of critical temperature observed in granular elemental superconductors \cite{deutscher1973granular, cohen1968superconductivity, cohen1967strong}.
The effect was interpreted as the surface being described by a different Hamiltonian, based on a conjecture of different chemical composition, which may indeed be the case, especially in complex compounds or with enhanced phonon interaction \cite{ginzburg1964surface, naugle1973evidence}.
However, higher critical temperature of the surface was reported also for clean elemental superconductors \cite{khlyustikov2011critical,khlyustikov2016surface}. 
Recently the claim of direct evidence for surface superconductivity was reported using the newly developed direct probe \cite{mangel2020stiffnessometer}.
Also, results in \cite{kapon2019phase} hint for possible interpretation in terms of surface critical temperature, which indeed should, in general, depend on the orientation of the boundary relative to the crystal axes.

The results from microscopic calculations and experiments show that the superconductor-insulator interface is nontrivial.
This is important for various applications.
For example, boundaries play a big role in quantum devices such as superconductors-based qubits and single-photon detectors, see, e.g. \cite{bommer2019spin,baghdadi2020enhancing}.
Moreover, the GL model remains the only nonlinear model amenable to the numerical solution at a substantially large length scale, required for modeling such devices.
The latter provides additional motivation for this work to revise the derivation of boundary conditions in the GL model.
Additionally, we resolve the ambiguity in boundary conditions when terms of higher-order in derivatives are added, see the discussion in \cite{samoilenka2020pair}.

This paper is organized in the following way:
In \secref{sec_derivation} we set up a microscopic BCS model, which is used to derive the GL model.
In \secref{sec_small_q} we derive that boundary conditions can be found from the mirror reflecting the order parameter in the boundary.
The surface term is neglected.
In \secref{sec_big_q} we microscopically obtain surface term $F_{surf}$ for several models.
The result is used in subsequent sections.
In \secref{sec_GL_h} we solve for a phase diagram that includes boundary states in the GL model for the spin imbalanced system.
In \secref{sec_GL} we obtain the difference of bulk and boundary critical temperatures in the GL model \eqref{F_GL}.
In \secref{sec_GL_H} we introduce the magnetic field and give a microscopic assessment of how $\gamma < 0$ enhances the third critical magnetic field $H_{c3}$.

\section{The microscopic model}\label{sec_derivation}
Consider continuous-space fermionic theory with the BCS type local attractive interaction given by strength $V > 0$.
We regularize the interaction by the Debye frequency $\omega_D$ such that only electrons with Matsubara frequency $< \omega_D$ interact.
The path integral formulation of this model is given by the action $S$ and the partition function $Z$ (see Chapter 6.4 in \cite{altlandsimonsBook}):
\begin{equation}\label{FermiHubbard}
\begin{gathered}
S = \int_0^{\frac{1}{T}} d\tau \int_{- \infty}^{+ \infty} d\textbf{r} \left[ \sum_{\sigma = \downarrow, \uparrow} a^\dagger_{\sigma} (\partial_\tau + \varepsilon_\sigma) a_{\sigma}
- V a^\dagger_{\uparrow} a^\dagger_{\downarrow} a_{\downarrow} a_{\uparrow} \right] \\
Z = \int D[a^\dagger, a] e^{-S}
\end{gathered}
\end{equation}

where $a_{\sigma}(\tau, \textbf{r}),\ a^\dagger_{\sigma}(\tau, \textbf{r})$ are Grassmann fields that correspond to fermionic creation and annihilation operators and depend on imaginary time $\tau$, $d$ dimensional space coordinates $\textbf{r}$ and spin $\sigma$.
Next, $\varepsilon_\sigma \equiv E - \mu_\sigma$, where $\mu_\sigma$ is the chemical potential, and $T$ is the temperature.
The single-electron energy is $E \equiv E(\ii \nabla)$ with $E(0) = 0$, which is $E(k) = \frac{k^2}{2 m}$ for free electrons.
It is assumed that $E$ depends only on the modulus of $\textbf{k}$ so that $E(\textbf{k}) \equiv E(|\textbf{k}|)$.
We consider a superconductor positioned in the $\Omega$ domain and an ideal insulator positioned everywhere else.
To model the insulator we assume that $\mu_\sigma$ is finite in $\Omega$ and $\mu_\sigma \to - \infty$ elsewhere.

We perform a Hubbard-Stratonovich transformation in the Cooper channel by introducing an auxiliary bosonic field $\Delta(\tau, \textbf{r})$:
\begin{equation}
\begin{gathered}
e^{ V \int d\tau d\textbf{r} a^\dagger_{\uparrow} a^\dagger_{\downarrow} a_{\downarrow} a_{\uparrow} } = \\
\int D[\Delta^\dagger, \Delta] e^{ - \int d\tau d\textbf{r} \left[ \frac{\Delta^\dagger \Delta}{V} + \Delta^\dagger a_{\downarrow} a_{\uparrow} + \Delta a^\dagger_{\uparrow} a^\dagger_{\downarrow} \right] }
\end{gathered}
\end{equation}

By introducing Nambu spinors $A^\dagger = \left( a^\dagger_{\uparrow}, a_{\downarrow} \right)$ and $A = 
\begin{pmatrix}
a_{\uparrow}\\
a^\dagger_{\downarrow}
\end{pmatrix}
$ we rewrite the partition function as
\begin{equation}
\begin{gathered}
Z = \int D[A^\dagger, A] D[\Delta^\dagger, \Delta] e^{- \int d\tau \int d\textbf{r} \left[ \frac{\Delta^\dagger \Delta}{V} + A^\dagger (\partial_\tau + H) A \right]} \\
H = 
\begin{pmatrix}
\varepsilon_{\uparrow} & \Delta \\
\Delta^\dagger & - \varepsilon_{\downarrow}
\end{pmatrix}
\end{gathered}
\end{equation}

Then, by performing the Berezin integral, we integrate out the fermionic degrees of freedom:
\begin{equation}\label{ZF_init}
\begin{gathered}
Z = \int D[\Delta^\dagger, \Delta] e^{- F / T} \\
F = - T \ln \det (\partial_\tau + H) + T \int d\tau \int d\textbf{r} \frac{\Delta^\dagger \Delta}{V}
\end{gathered}
\end{equation}

where $F$ is the free energy.
Next, we make mean-field assumptions: that $\Delta$ is a classical field (does not depend on $\tau$) and that it does not fluctuate thermally and is found as a minimum of $F$.
In this approximation the problem simplifies:
\begin{equation}\label{lndet}
\begin{gathered}
\ln{\det\left( \partial_\tau + H \right)}
= \text{Tr} \sum_{n}^{|\omega_n| < \omega_D} \ln\left( \ii \omega_n + H \right) \\
= \text{Tr} \sum_{n} \ln\left( 1 + (\ii \omega_n + H_0)^{-1} \Lambda \right)\\
= - \sum_{n} \sum_{k = 1}^\infty \frac{(-1)^k}{k} \text{Tr}\left[ \left( G_\uparrow \Delta G_\downarrow^* \Delta^* \right)^k \right]
\end{gathered}
\end{equation}

where Matsubara frequencies are $\omega_n = 2 \pi T \left( n + 1/2 \right)$ and we used
\begin{equation}
\begin{gathered}
H_0 = 
\begin{pmatrix}
\varepsilon_{\uparrow} & 0 \\
0 & - \varepsilon_{\downarrow}
\end{pmatrix},\ \ 
\Lambda = 
\begin{pmatrix}
0 & \Delta \\
\Delta^\dagger & 0
\end{pmatrix}\\
(\ii \omega_n + H_0)^{-1}(\textbf{r}, \textbf{r}') =
\begin{pmatrix}
G_\uparrow(\textbf{r}, \textbf{r}') & 0 \\
0 & - G_\downarrow^*(\textbf{r}, \textbf{r}')
\end{pmatrix}
\end{gathered} 
\end{equation}

Green's functions for spin $\sigma$ electrons are determined from:
\begin{equation}
\begin{gathered}
(\ii \omega_n + \varepsilon_\sigma) G_\sigma(\textbf{r}, \textbf{r}') = \delta(\textbf{r} - \textbf{r}')
\end{gathered} 
\end{equation}

Since $\mu_\sigma \to - \infty$ in the insulator, the single-electron wave functions will be zero there.
This results in boundary conditions for Green's functions in the following way: for coordinate $\textbf{r}_b$ lying on the boundary $\partial \Omega$ we get
\begin{equation}\label{G_bc}
G_\sigma(\textbf{r}, \textbf{r}_b) = G_\sigma(\textbf{r}_b, \textbf{r}) = 0
\end{equation}

Now let us consider $\Omega : x > 0$.
In that case the Green's function can be obtained in the form:
\begin{equation}\label{G_def}
\begin{gathered}
G_\sigma(\textbf{r}, \textbf{r}') = g_\sigma(\textbf{r} - \textbf{r}') - g_\sigma(\textbf{r} - \underline{\textbf{r}}') \\
g_\sigma(\textbf{r}) = \frac{1}{(2 \pi)^d} \int_{- \infty}^{+ \infty} \frac{e^{\ii \textbf{k} \textbf{r}}}{\ii \omega_n + \varepsilon_\sigma(k)} d\textbf{k}
\end{gathered}
\end{equation}

where $g_\sigma(\textbf{r})$ is bulk Green's function, $\underline{\textbf{r}} \equiv \textbf{r} - 2 x \hat{\textbf{x}}$ and $k \equiv |\textbf{k}|$.

Below we assume the following:
\begin{equation}\label{inequalities}
\mu_\sigma \gg \omega_D \gg T_c \gg Q v_F
\end{equation}

where $T_c$ is the critical temperature, $v_F$ is the Fermi velocity, and we are interested in a slow-varying order parameter with momentum $|\textbf{q}| \lesssim Q$ (this justifies the expansion in $q$ that we do later).
Close to the transition to the normal state, $\Delta \to 0$.
Hence we can truncate the expansion in $\Delta$ in \eqref{lndet}.
For usual superconductors, $\mu_\uparrow = \mu_\downarrow$.
We also consider a superconductor with a spin imbalance.
There, GL expansion is done near the tricritical point associated with the bulk Fulde-Ferrel-Larkin-Ovchinnikov (FFLO) state \cite{FuldeFerrell1964,LarkinOvchinnikov1964} (in which case $|\mu_\uparrow - \mu_\downarrow|$ is of the order of $T_c$). 

Combining \eqref{ZF_init} and \eqref{lndet} we obtain the quadratic-in-$\Delta$ part of the free energy:
\begin{equation}\label{F2_Omega}
\begin{gathered}
F_2 = \int_{\Omega} d\textbf{r} \frac{|\Delta|^2}{V} \\
- T \sum_{n}^{|\omega_n| < \omega_D} \int_{\Omega} d\textbf{r} d\textbf{r}' G_\uparrow(\textbf{r}, \textbf{r}') \Delta(\textbf{r}') G_\downarrow^*(\textbf{r}', \textbf{r}) \Delta^*(\textbf{r})
\end{gathered}
\end{equation}

To simplify $F_2$ we need to perform Fourier transform.
However, integrals over $x$ in $F_2$ are over half space.
To extend them to full space we note that
\begin{equation}\label{G_reflection}
G_\sigma(\underline{\textbf{r}}, \textbf{r}') = G_\sigma(\textbf{r}, \underline{\textbf{r}}') = - G_\sigma(\textbf{r}, \textbf{r}')
\end{equation}

Together with defining $\Delta(\underline{\textbf{r}}) = \Delta(\textbf{r})$, without loss of generality we obtain:
\begin{equation}\label{F2_inf}
\begin{gathered}
F_2 = \frac{1}{2} \int_{- \infty}^{+ \infty} d\textbf{r} \frac{|\Delta|^2}{V} \\
- \frac{T}{4} \sum_{n} \int_{- \infty}^{+ \infty} d\textbf{r} d\textbf{r}' G_\uparrow(\textbf{r}, \textbf{r}') \Delta(\textbf{r}') G_\downarrow^*(\textbf{r}', \textbf{r}) \Delta^*(\textbf{r})
\end{gathered}
\end{equation}

\begin{figure}
\centering
\includegraphics[width=0.99\linewidth]{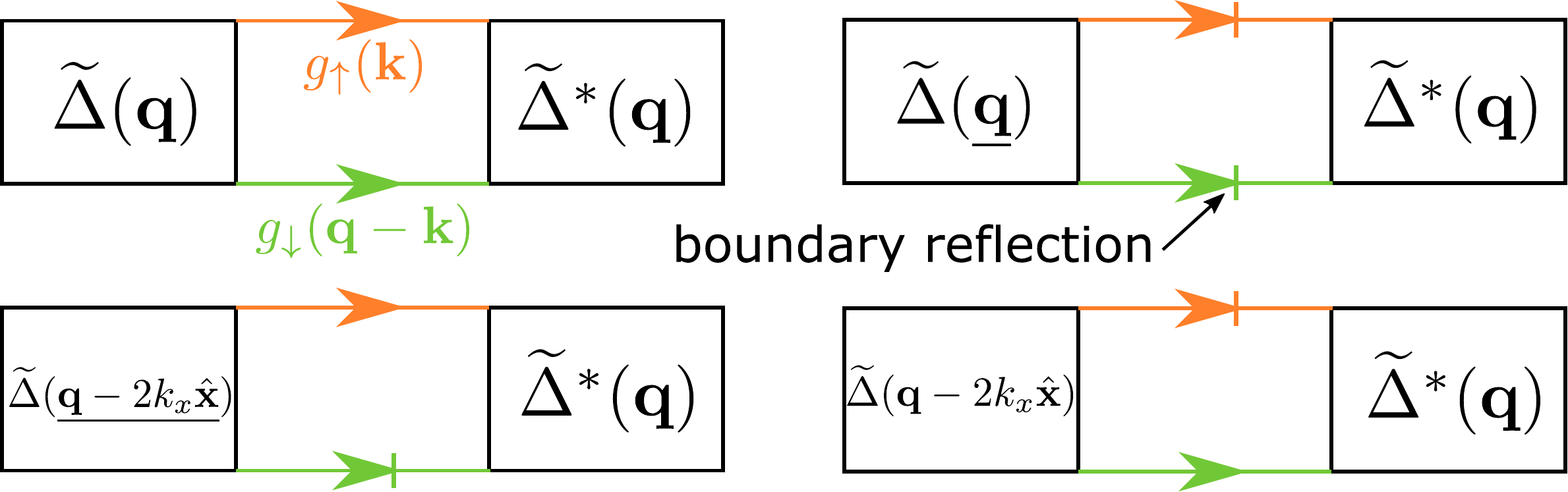}
\caption{
Illustrations of free-energy terms which are second-order in the order parameter \eqref{F2_fourier}.
Orange (green) lines denote bulk Green's functions of spin-up (spin-down) free electrons (reflection from the boundary is marked by a vertical line).
Squares denote the order parameter $\Delta$.
In all cases, the particle-particle bubble, consisting of two Green's functions, is explicitly given by $f(k, \textbf{k} - \textbf{q})$, see \eqref{f_D_def}.
The top two diagrams with zero and two boundary reflections lead to usual bulk-like terms, where order parameters with the same momentum are coupled.
They give terms proportional to $D(\textbf{q})$ in \eqref{F2_fourier}.
The bottom two diagrams have one reflected from the boundary electron.
The latter results in coupling between order parameters of different momenta (the last term in free energy \eqref{F2_fourier}).
As shown below, this gives rise to boundary states.
}
\label{fig_diagrams}
\end{figure}

Performing the Fourier transform using \eqref{G_def} and $\Delta(\textbf{r}) = \frac{1}{(2 \pi)^d} \int_{- \infty}^{+ \infty} \widetilde{\Delta}(\textbf{q}) e^{\ii \textbf{q} \textbf{r}} d\textbf{q}$ we obtain, see \figref{fig_diagrams}:
\begin{equation}\label{F2_fourier}
\begin{gathered}
F_2 = \frac{1}{2} \int_{- \infty}^{+ \infty} \frac{d\textbf{q}}{(2 \pi)^d} \left[ \left( \frac{1}{V} - D(\textbf{q}) \right) |\widetilde{\Delta}(\textbf{q})|^2 \right. \\
\left. + \int_{- \infty}^{+ \infty} \frac{d\textbf{k}}{(2 \pi)^d} \widetilde{\Delta}(\textbf{q} - 2 k_x \hat{\textbf{x}}) \widetilde{\Delta}^*(\textbf{q}) f(k, \textbf{k} - \textbf{q}) \right]
\end{gathered}
\end{equation}

where
\begin{equation}\label{f_D_def}
\begin{gathered}
f(k, \textbf{k} - \textbf{q}) = T \sum_n^{|\omega_n| < \omega_D} \frac{1}{\ii \omega_n + \varepsilon_\uparrow(k)} \frac{1}{- \ii \omega_n + \varepsilon_\downarrow(\textbf{k} - \textbf{q})} \\
D(\textbf{q}) = \int_{- \infty}^{+ \infty} \frac{d\textbf{k}}{(2 \pi)^d} f(k, \textbf{k} - \textbf{q})
\end{gathered}
\end{equation}

to simplify \eqref{F2_fourier} note that $f(k, \textbf{k} - \textbf{q})$ is not negligible\footnote{Explicitly for $\omega_D \gg T$ we obtain $f(k, k') \simeq \frac{\pi \left( \tanh\frac{\varepsilon_\uparrow}{2 T} + \tanh\frac{\varepsilon_\downarrow'}{2 T} \right) - 2 \left( \arctan\frac{\varepsilon_\uparrow}{\omega_D} + \arctan\frac{\varepsilon_\downarrow'}{\omega_D} \right)}{2 \pi \left( \varepsilon_\uparrow + \varepsilon_\downarrow' \right)}$. Note that, equivalently, we could have defined $\omega_D$ cutoff for energies $|\varepsilon| < \omega_D$ and let the sum over Matsubara frequencies be unrestricted.} only at $|\varepsilon_\uparrow(k)| \lesssim \omega_D$ and $|\varepsilon_\uparrow(\textbf{k} - \textbf{q})| \lesssim \omega_D$.
Since $|\mu_\uparrow - \mu_\downarrow| \ll \omega_D \ll \mu_\sigma$ we define $\mu = \frac{\mu_\uparrow + \mu_\downarrow}{2}$ and $h = \frac{\mu_\uparrow - \mu_\downarrow}{2}$, and hence $\varepsilon(k) = E(k) - \mu$.
Then the Fermi momentum $k_F: \varepsilon(k_F) = 0$ and the Debye momentum $k_D: |\varepsilon(k_F \pm k_D)| \lesssim \omega_D$, which can be estimated as $k_D \simeq \omega_D / v_F$, where the Fermi velocity $v_F = E'(k_F)$.
Hence, $f(k, \textbf{k} - \textbf{q})$ is nonzero when $k$ and $\textbf{k} - \textbf{q}$ are on a Fermi sphere of radius $k_F$ and thickness $2 k_D$.

\begin{figure}
\centering
\includegraphics[width=0.99\linewidth]{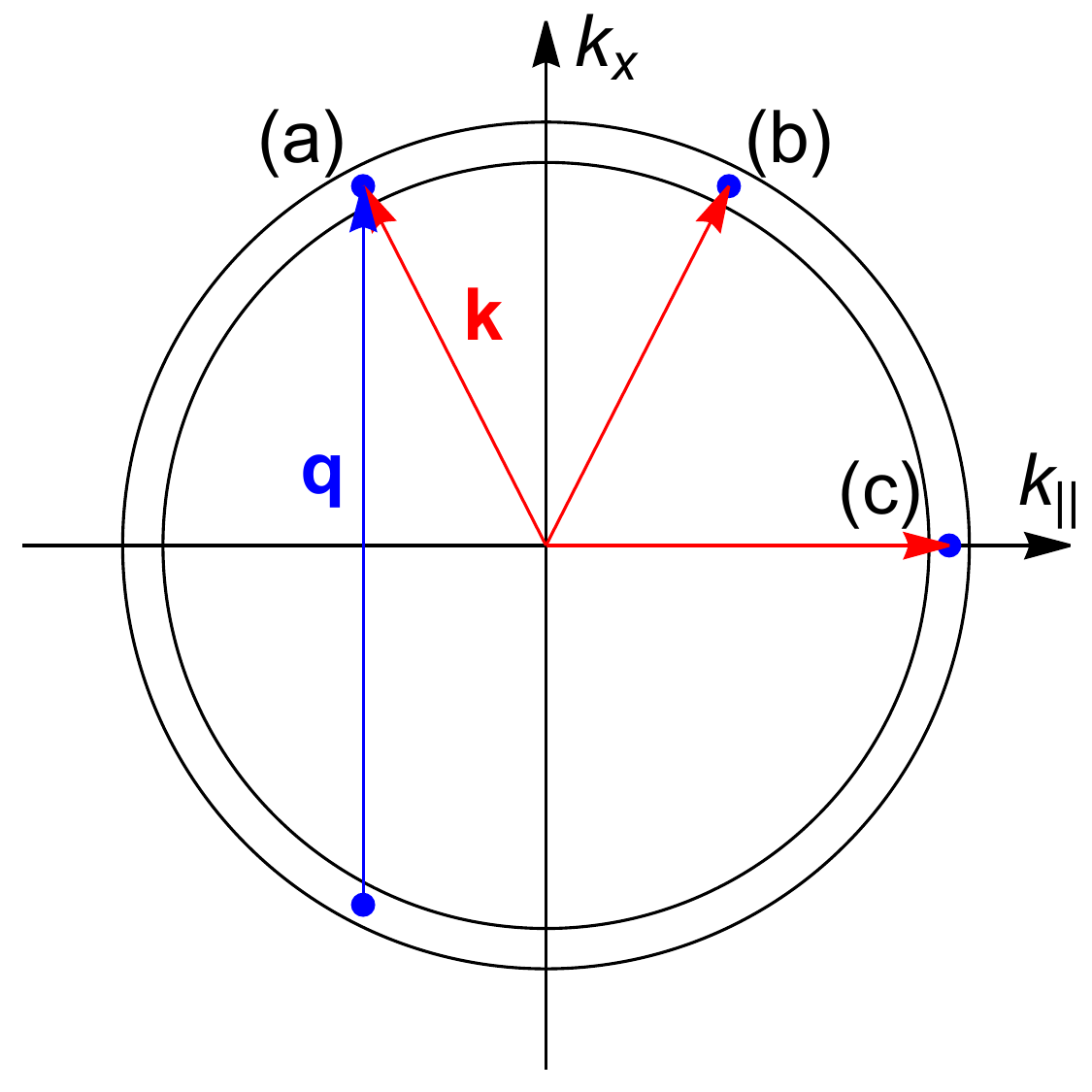}
\caption{
Fermi sphere and three types of configurations of momenta (a)-(c), that give rise to boundary states.
In all these cases, particle-particle bubble $f(k, \textbf{k} - \textbf{q})$ (defined by \eqref{f_D_def}) is large.
This is due to the fact that $\textbf{q}$ (blue) and $\textbf{k}$ (red) lie in the Fermi sphere (black circles).
In all cases $|\textbf{k}| - k_F \in [ - k_D, k_D]$.
By $k_x$ we denote momentum orthogonal to the boundary and by $k_{||}$ momentum parallel to it.
Configurations can be described by the following values of $k_x$ and $\textbf{q}$:
\textbf{(a)} $| k_x | \gg k_D$ and $\textbf{q}$ is large.
Namely, $| \textbf{q} - 2 k_x \hat{\textbf{x}} | \lesssim Q$.
\textbf{(b)} $| k_x | \gg k_D$ and $| \textbf{q} | \lesssim Q$.
\textbf{(c)} $| k_x | \lesssim k_D$ and $| \textbf{q} | \lesssim Q$.
Note, that, for two- and three-dimensional systems all three appear, whereas for one-dimensional systems, only cases (a) and (b) appear with $k_{||} = 0$.
}
\label{fig_fermi_sphere}
\end{figure}

Usually it is assumed that $\Delta$ varies slowly with momentum $q \lesssim Q \ll k_D$ \eqref{inequalities}.
However, this is true only for directions parallel to the boundary.
By contrast, in the $x$ direction, fast oscillations with $q \lesssim 2 k_F$ are present, the gap exhibits Friedel oscillations \cite{giamarchi1990onset,samoilenka2020boundary}. 
The existence of oscillations, means that $f(k, \textbf{k} - \textbf{q})$ has contributions from three different types of points on Fermi sphere illustrated in \figref{fig_fermi_sphere}.

In this work, we are interested in the description of the boundary of a superconductor at the level of the GL model.
In the GL approximation, the order-parameter field is coarse-grained and thus varies slowly in real space, which corresponds to small momentum $|\textbf{q}| \lesssim Q$, below it is denoted by $\widetilde{\psi}(\textbf{q}) \equiv \widetilde{\Delta}(\textbf{q})$.
Whereas the order parameter that changes fast in the $x$ direction (large momenta $Q \lesssim |q_x| \lesssim 2 k_F$) we denote $\widetilde{\psi}_f(\textbf{q}) \equiv \widetilde{\Delta}(\textbf{q})$.

\section{Small-q contribution to boundary conditions: de Gennes approximation}\label{sec_small_q}
We begin by reproducing the de Gennes microscopic boundary conditions under an approximation similar to that used in \cite{abrikosov1965concerning,deGennes_Boundary,CdGM_french,book_de_Gennes} considering the small-$q$ contribution to \eqref{F2_fourier}.
That is, in this section, we consider only the contribution from a slowly varying order parameter $\psi$.
In this approximation we restrict $|\textbf{q}| < Q$ in $\widetilde{\Delta}(\textbf{q})$.
Then we obtain
\begin{equation}
F_2^0 \simeq \frac{1}{2} \int_{- Q}^{+ Q} \frac{d\textbf{q}}{(2 \pi)^d} \left( \frac{1}{V} - D(\textbf{q}) \right) |\widetilde{\psi}(\textbf{q})|^2
\end{equation}

Since $\textbf{q}$ is small in \eqref{inequalities} we expand $D(\textbf{q})$ as:
\begin{equation}\label{F20_q}
F_2^0 \simeq \int_{- Q}^{+ Q} \frac{d\textbf{q}}{(2 \pi)^{d}} |\widetilde{\psi}(\textbf{q})|^2 \left[ c_0 + c_2 q^2 + c_4 q^4 + .. \right]
\end{equation}

where $c_i$ are the usual bulk coefficients \cite{Abrikosov_Gorkov_Dzyaloshinski_Book}.
Here we carry out the calculation in a more general form that is also applicable for the case where one has to keep higher-order derivative terms, such as the case of spin imbalanced superconductors, including those in the FFLO state \cite{buzdin1997generalized,Radzihovsky2011}.
Transforming \eqref{F20_q} into real space gives:
\begin{equation}\label{F20_r}
F_{2G}^{0} \simeq \int_{- \infty}^{+ \infty} d\textbf{r} \left[ c_0 |\psi|^2 + c_2 |\nabla \psi|^2 + c_4 |\nabla^2 \psi|^2 + .. \right]
\end{equation}

where terms with derivatives can be written in \textit{any} integrated-by-parts form like $- c_2 \psi \nabla^2 \psi^* + c_2 \psi \nabla^4 \psi^* + ..$.
This is possible since integral over $x$ in \eqref{F20_r} is over $(- \infty, + \infty)$ with $\psi$ reflected in the boundary as $\psi(\underline{\textbf{r}}) = \psi(\textbf{r})$.
Hence when going back to the actual system $\Omega: x > 0$ one obtains boundary conditions at $x = 0$, which depend on the chosen form of $c_i$ terms.

For example, if the $c_4$ term is not included, the $c_2$ term gives the usual \cite{deGennes_Boundary,CdGM_french,book_de_Gennes,abrikosov1965concerning}
\begin{equation}\label{bc_0}
\textbf{n} \cdot \nabla \psi = 0
\end{equation}
This is the de Gennes boundary condition. Applying it dictates that the critical temperature of bulk and boundaries are identical.
To derive it, note that $\psi$ is continuous across $x = 0$ but not necessarily smooth.
Hence, writing $c_2 |\nabla \psi|^2$ has no additional terms at the surface and boundary conditions \eqref{bc_0} are obtained by variation as usual.
Alternatively we could have picked $- c_2 \psi^* \nabla^2 \psi$.
In that case
\begin{equation}\label{ddpsi}
\nabla^2 \psi = 2 \delta(x) \left. \partial_x \psi\right|_{x \to 0^+} + \left. \nabla^2 \psi \right|_{x > 0}
\end{equation}
Integrating by parts we obtain again $c_2 |\nabla \psi|^2$, since the $delta$ function in \eqref{ddpsi} compensates for integration by parts.

Consider now the case where the $c_4$ term is included, such as, for example, in the GL model of a spin imbalanced uniform and FFLO systems \cite{buzdin1997generalized,Radzihovsky2011}.
Then $c_4 |\nabla^2 \psi|^2$ with \eqref{ddpsi} shows that in order for energy to be finite, we need first of the two boundary conditions: 
\begin{equation}\label{bc_h_0}
\textbf{n} \cdot \nabla \psi = 0,\ \ \ \textbf{n} \cdot \nabla^3 \psi = 0
\end{equation}
where the second boundary condition is obtained from variation of the energy.

This provides microscopically derived boundary conditions in de Gennes approximation for moderately spin-imbalanced systems and FFLO systems.
Boundary conditions \eqref{bc_h_0} were used before for these systems, see e.g. \cite{samokhin2019FFLO,plastovets2020onions}.
Note that these boundary conditions eliminate the PDW boundary states discussed in \cite{barkman2019surface,samoilenka2020pair} and contradict the phenomenological GL boundary conditions used there.
At the same time, the PDW boundary states are unambiguously demonstrated in the full microscopic model \cite{samoilenka2020pair}.
We resolve that question below.

Note, that this boundary condition of the order parameter being reflected, $\Delta(\underline{\textbf{r}}) = \Delta(\textbf{r})$, follows from the general property of the Green's function \eqref{G_reflection}.
Hence it is easy to generalize it to different systems.
For example, in noncentrosymmetric superconductors \cite{book_noncentro_sym} with local interaction \eqref{FermiHubbard} the analog of \eqref{F2_Omega} just has a matrix $G$ and $\Delta$ \cite{samoilenka2020spiral}, while the property \eqref{G_reflection} follows from the boundary condition for the Green's function \eqref{G_bc}.
Hence the counterpart of de Gennes boundary conditions, in that case, is obtained from reflecting the fields at the boundary as well.
For the simplest model they are equivalent to boundary conditions obtained from variation of the free-energy function, see \cite{samoilenka2020spiral}. 

To summarize this section: we reported generalized to GL models with higher-derivatives derivation of de Gennes boundary conditions.
However, these conditions do not reproduce the superconducting boundary states.
On the other hand, it was shown microscopically that these states exist on the macroscopic length scale \cite{samoilenka2020pair} and hence they should be reproducible in microscopically derived GL models.
This problem is resolved in the next section.

\section{The boundary conditions beyond the de Gennes approximation}\label{sec_big_q}
In this section, we consider whether there is a nonvanishing contribution from the terms in \eqref{F2_fourier} coming from an averaging of the fast-oscillating order parameter.
We denote them $F_2^1$ such that $F_2 = F_2^0 + F_2^1$.
These terms have a large-$\textbf{q}$ counterpart of $F_2^0$ \eqref{F20_q} and $(a)$, $(b)$, $(c)$ parts of $F_2$, see \figref{fig_fermi_sphere}:
\begin{equation}\label{F21_q}
\begin{gathered}
F_2^1 \simeq \frac{1}{2} \int_{- Q}^{+ Q} \frac{d\textbf{q}_{||}}{(2 \pi)^{d - 1}} \int' \frac{dq_x}{2 \pi} \left( \frac{1}{V} - D(\textbf{q}) \right) |\widetilde{\psi}_f(\textbf{q})|^2 \\
+ \frac{1}{2} \int_{- Q}^{+ Q} \frac{d\textbf{q}}{(2 \pi)^d} \left[ \int_{- Q'}^{+ Q'} \frac{dk_x}{2 \pi} \widetilde{\psi}(\textbf{q} - 2 k_x \hat{\textbf{x}}) \widetilde{\psi}^*(\textbf{q}) f_1(k_x, k_x - q_x) \right. \\
\left. + \int' \frac{dk_x}{2 \pi} \widetilde{\psi}_f(\textbf{q} - 2 k_x \hat{\textbf{x}}) \widetilde{\psi}^*(\textbf{q}) f_1(k_x, k_x) \right. \\
\left. + \int' \frac{dk_x}{2 \pi} \widetilde{\psi}(\textbf{q}) \widetilde{\psi}_f^*(\textbf{q} - 2 k_x \hat{\textbf{x}}) f_1(k_x, k_x) \right]
\end{gathered}
\end{equation}

where $Q' \gtrsim Q$ and $\textbf{q}_{||}$ are $\textbf{q}$ components parallel to the boundary if there are any.
By $\int' dp$ we denote integral $\int_{- \infty}^{+ \infty} dp$ excluding $|p| \lesssim k_D$.
We defined:
\begin{equation}\label{f1_def}
f_1(q, p) = \int_{- \infty}^{+ \infty} \frac{d\textbf{k}_{||}}{(2 \pi)^{d - 1}} f\left(\sqrt{q^2 + k_{||}^2}, \sqrt{p^2 + k_{||}^2}\right)
\end{equation}

\eqref{F21_q} is simplified to:
\begin{equation}\label{F21_r}
\begin{gathered}
F_2^1 \simeq \int_{- \infty}^{+ \infty} d \textbf{r}_{||} \int' \frac{dp}{2 \pi} \left[ \left( \frac{1}{V} - D(2 p) \right) |\widetilde{\psi}_f(2 p, \textbf{r}_{||})|^2 \right. \\
\left. + \frac{1}{2} \left( \widetilde{\psi}_f(2 p, \textbf{r}_{||}) \psi^*(0, \textbf{r}_{||}) + \psi(0, \textbf{r}_{||}) \widetilde{\psi}_f^*(2 p, \textbf{r}_{||}) \right) f_1(p, p) \right] \\
+ \frac{f_1(0, 0)}{4} \int_{- \infty}^{+ \infty} d \textbf{r}_{||} |\psi(0, \textbf{r}_{||})|^2
\end{gathered}
\end{equation}

where we used $\int_{- Q}^{+ Q} \frac{d q_x}{2 \pi} \widetilde{\psi}(q_x, \textbf{r}_{||}) = \psi(0, \textbf{r}_{||})$.
Note, that the full field $\Delta(x)$ is zero at the boundary since electrons are perfectly reflected from it and hence Green's functions are zero there.
However, the GL order parameter $\psi$ represents only a slowly varying part of the pairing field, which can be nonzero at the boundary.

By varying \eqref{F21_r} with respect to $\widetilde{\psi}_f$ we obtain the solution for fast-oscillating part of the order parameter: 
\begin{equation}
\widetilde{\psi}_f(2 p, \textbf{r}_{||}) = - \frac{\psi(0, \textbf{r}_{||})}{2} \frac{f_1(p, p)}{\frac{1}{V} - D(2 p)}
\end{equation}

Inserting it back into \eqref{F21_r} we get the surface term:
\begin{equation}\label{gamma_res}
\begin{gathered}
F_2^1 = \gamma \int_{- \infty}^{+ \infty} d \textbf{r}_{||} |\psi(0, \textbf{r}_{||})|^2 \\
\gamma = - \frac{1}{4} \left[ \int' \frac{dp}{2 \pi} \frac{f_1^2(p, p)}{\frac{1}{V} - D(2 p)} - f_1(0, 0) \right]
\end{gathered}
\end{equation}

Now let us analyze that contribution in various dimensions.

\subsection{Superconducting wire}
For one-dimensional systems, we have no contribution associated with configuration $(c)$, see \figref{fig_fermi_sphere}, in energy \eqref{F21_q} and hence there should be no $f_1(0, 0)$ term in the expression for the boundary term \eqref{gamma_res}.
However we can use the same formula \eqref{gamma_res} since for $d = 1$ we have $f_1(p, q) = f(p, q)$ and hence $f_1(0, 0) = 0$.
We simplify \eqref{gamma_res} as:
\begin{equation}
\gamma \simeq - \frac{1}{2} \int_{k_F - k_D}^{k_F + k_D} \frac{dp}{2 \pi} \frac{f^2(p, p)}{\frac{1}{V} - D(2 p)}
\end{equation}
and estimate $D(2 p) \simeq \frac{N}{2}$ for $|p - k_F| \lesssim k_D$, where the density of states at the Fermi level is $N = \frac{1}{\pi v_F}$.
By performing the integral over $p$ we obtain
\begin{equation}\label{gamma_1d_h}
\gamma \simeq - \frac{N V}{1 - \frac{N V}{2}} \frac{2 \pi \coth(\frac{h}{T}) \text{Im}\Psi^{(1)}\left(Z\right) - \text{Re}\Psi^{(2)}\left(Z\right)}{32 \pi^2 T}
\end{equation}

where $Z = \frac{1}{2} - \ii \frac{h}{2 \pi T}$ and $\Psi^{(n)}$ are polygamma functions of order $n$.
For $h = 0$ \eqref{gamma_1d_h} reduces to:
\begin{equation}\label{gamma_1d}
\gamma \simeq - \frac{N V}{1 - \frac{N V}{2}} \frac{7 \zeta(3)}{8 \pi^2 T}
\end{equation}
where $\zeta$ is the Riemann zeta function.

Therefore, in one-dimensional GL theory, there is a boundary term for the interface between a superconductor and a vacuum \eqref{gamma_res}.
The term has a negative microscopically derived prefactor $\gamma$.
This implies that the gap is increased near the boundary and there are superconducting boundary states \cite{samoilenka2020boundary}.
The conclusion applies both to quasifree and band fermions.

\subsection{Planar superconductor}
Consider the one-dimensional boundary of a two-dimensional sample.
We estimate $f_1(p, p)$ for $k_F - k_D \gtrsim |p|$ as:
\begin{equation}\label{f1_2d}
f_1(p, p) \simeq \frac{2 N L}{k_F} \frac{1}{\sqrt{1 - \frac{p^2}{k_F^2}}}
\end{equation}

with
\begin{equation}\label{L_def}
L = \ln \frac{\omega_D}{2 \pi T} - \text{Re}\Psi(Z) = \frac{1}{N V} - \frac{\alpha}{N}
\end{equation}

where $\alpha$ is defined in \eqref{parameters_h}, $T_{c1} = \frac{2 e^{\gamma_E}}{\pi} \omega_D e^{- \frac{1}{N V}}$ is the bulk critical temperature, $\gamma_E$ is Euler gamma, $\Psi$ is digamma function and $N$ is density of states at Fermi level.
In this work, we consider the case of $d$ dimensional system in the BCS limit where $N V$ is small.
For the two dimensional case, $N = \frac{k_F}{2 \pi v_f}$.
The function $D(2 p)$ for $k_F - k_D \gtrsim |p| \gtrsim k_D$ is then given by:
\begin{equation}\label{D_2d}
D(2 p) \simeq \frac{2 \ln 2}{\pi} N \frac{k_D}{|p|} \frac{1}{\sqrt{1 - \frac{p^2}{k_F^2}}}
\end{equation}

\eqref{f1_2d} and \eqref{D_2d} allow us to compute $\gamma$ in \eqref{gamma_res} up to logarithmic accuracy:
\begin{equation}\label{gamma_2d}
\gamma \simeq - \frac{\ln \frac{k_F}{k_D}}{2 \pi k_F V}
\end{equation}

Therefore, similarly to the one-dimensional case of almost free fermions, we recover the boundary states at the level of the GL theory.

\subsection{Three dimensional isotropic sample}
Next we consider the two-dimensional boundary of a three-dimensional sample.
In this case $f_1$ can be estimated as:
\begin{equation}\label{f1_3d}
f_1(p, p) \simeq
\begin{cases}
|p| \lesssim k_F,\ \frac{\pi}{k_F} N L \\
|p| \gtrsim k_F,\ 0
\end{cases}
\end{equation}
The density of states of a three dimensional superconductoris $N = \frac{k_F^2}{2 \pi^2 v_F}$.
Whereas $D(2 p)$ for $k_D \lesssim |p| \lesssim k_F$ is
\begin{equation}
D(2 p) \simeq \frac{k_D}{|p|} N \ln 2
\end{equation}
and $D(2 p) \simeq 0$ for $|p| \gtrsim k_F$.
This allows us to calculate the integral in \eqref{gamma_res}, which gives:
\begin{equation}\label{gamma_3d}
\gamma \simeq \frac{\pi N L}{4 k_F} \left( V \alpha + \frac{k_D}{k_F} \left( c - N V \ln 2 \ln \frac{k_F}{c k_D} \right) \right).
\end{equation}
Here $c$ is the cut off parameter of order one defined by $\int' dp = 2 \int_{c k_D}^{\infty} dp$.
\eqref{gamma_3d} shows that we cannot justifiably calculate $\gamma$ using this analytical approach since it depends on $c$.
Namely, as seen from \eqref{gamma_res}, in our approximation we can obtain $\gamma$ only of order $f_1(0, 0)$ or larger, whereas the leading-order contribution to the integral gives: $\int' \frac{dp}{2 \pi} \frac{f_1^2(p, p)}{\frac{1}{V} - D(2 p)} \simeq f_1(0,0)$.
Hence, this level of approximation indicates that $\gamma$ is very small for this model: of order \eqref{gamma_3d} or smaller, or could be zero. 
This is the property of the special case: a two-dimensional boundary of an isotropic continuous BCS model in three dimensions.
Note, that the situation is different in the three dimensional tight-binding BCS model \cite{samoilenka2020boundary}.

\subsection{The boundary states and anisotropy in three dimensions}
Since many of the superconducting materials of current interest are strongly anisotropic, let us explicitly consider the effects of anisotropy.
Consider the three-dimensional model that has single electron energy $E\left( \sqrt{(k_x / a)^2 + \textbf{k}_{||}^2} \right)$, which is anisotropic for $a \neq 1$.
Then $\gamma$ is given by the same formula \eqref{gamma_res} with the replacement $p \to p / a$ inside the single electron energies $E$.
Hence, using \eqref{f1_3d} we obtain the leading-order estimate:
\begin{equation}\label{gamma_3d_a}
\gamma \simeq - \frac{\pi (a - 1)}{4 k_F V}
\end{equation}

where $k_F$ is the Fermi momentum parallel to the boundary.
Hence, for $a > 1$, the superconductivity is enhanced near the boundary and there are boundary states.
In contrast, the gap is suppressed for $a < 1$.
In other words, boundary states are present if the Fermi sphere is stretched in the direction orthogonal to the boundary.
Note that \cite{kapon2019phase} reported a difference in $T_c$ for samples with different orientations of the surfaces relative to crystal axes using the stiffnessometer experiment.
The stifnessometer setup was proposed to resolve surface superconductivity \cite{mangel2020stiffnessometer}.

\section{Boundary states in superconductor with imbalanced fermions}\label{sec_GL_h}

Now we consider the case where superconducting pairing takes place in a model with spin imbalance, i.e., unequal densities of spin components.
In an infinite system, when there is a critical disparity of Fermi momenta of spin-up and spin-down fermionic components, the system undergoes a phase transition into an inhomogeneous FFLO state \cite{FuldeFerrell1964,LarkinOvchinnikov1964}. 
In such a state, the system has a modulation in the phase or modulus of the order-parameter field.
At the level of the Ginzburg-Landau theory, a phase transition into such a state manifests itself through the coefficient in front of the quadratic gradient term becoming negative.
Therefore, for the energy to be bounded from below, one needs to retain higher-order gradient terms with positive prefactors.

Combining the results of the microscopic derivation for bulk \cite{buzdin1997generalized} and boundary \secref{sec_derivation} we obtain the GL model with spin imbalance:
\begin{equation}\label{F_GL_h}
\begin{gathered}
F = \int_\Omega d\textbf{r} \left[ \alpha |\psi|^2 + K |\nabla \psi|^2 + \beta |\psi|^4 + \widetilde{K} |\nabla^2 \psi|^2 \right. \\
\left. + K_1 |\psi \nabla\psi|^2 + \frac{K_1}{8} \left( (\psi^* \nabla \psi)^2 + (\psi \nabla \psi^*)^2 \right) + \nu |\psi|^6 \right] \\
+ \gamma \int_{\partial \Omega} d\textbf{r}_{||} |\psi|^2
\end{gathered}
\end{equation}

Let us now study the problem of the boundary conditions in the presence of the higher-order derivative terms.
These conditions are obtained from considering the mirror reflected model.
The condition of finiteness of energy and variation of $F$ with respect to $\psi^*$ gives us the boundary conditions:
\begin{equation}\label{bc_h}
\textbf{n} \cdot \nabla \psi = 0,\ \ \ \textbf{n} \cdot \nabla^3 \psi = \frac{\gamma}{\widetilde{K}} \psi
\end{equation}
where $\textbf{n}$ is a unit vector pointing outside of the sample.
Note, that boundary conditions \eqref{bc_h} differ from the phenomenological boundary conditions used in \cite{barkman2019surface,samoilenka2020pair}.
Whereas the de Gennes boundary conditions, previously used for FFLO systems \cite{samokhin2019FFLO,plastovets2020onions} correspond to setting $\gamma = 0$ in our boundary conditions \eqref{bc_h}.
We derive that $\gamma$ is not zero.
In \eqref{F_GL_h} and \eqref{bc_h} $\gamma$ is given by \eqref{gamma_res} while the other parameters are \cite{buzdin1997generalized}:
\begin{equation}\label{parameters_h}
\begin{gathered}
\alpha = N \left[ \ln \frac{T}{T_{c1}} + \text{Re}\Psi(Z) - \Psi(1 / 2) \right] \\
\beta = - \frac{N \text{Re}\Psi^{(2)}(Z)}{2 (4 \pi T)^2},\ \ \ \nu = \frac{N \text{Re}\Psi^{(4)}(Z)}{12 (4 \pi T)^4} \\
K = \frac{v_F^2 \beta}{d},\ \ \widetilde{K} = \frac{3 v_F^4 \nu}{2 d (d + 2)},\ \ K_1 = \frac{4 v_F^2 \nu}{d}
\end{gathered}
\end{equation}

Let us consider now the boundary physics of a spin-imbalanced superconductor.
First, we determine when the superconductor transitions to a normal state.
We assume that this transition is of second order and hence $\psi \to 0$ at the transition.
Hence it is sufficient to solve linearized GL equations that follow from the variation of \eqref{F_GL_h}:
\begin{equation}\label{GL_eq_h_lin}
\alpha \psi - K \nabla^2 \psi + \widetilde{K} \nabla^4 \psi = 0
\end{equation}

\begin{figure}
\centering
\includegraphics[width=0.99\linewidth]{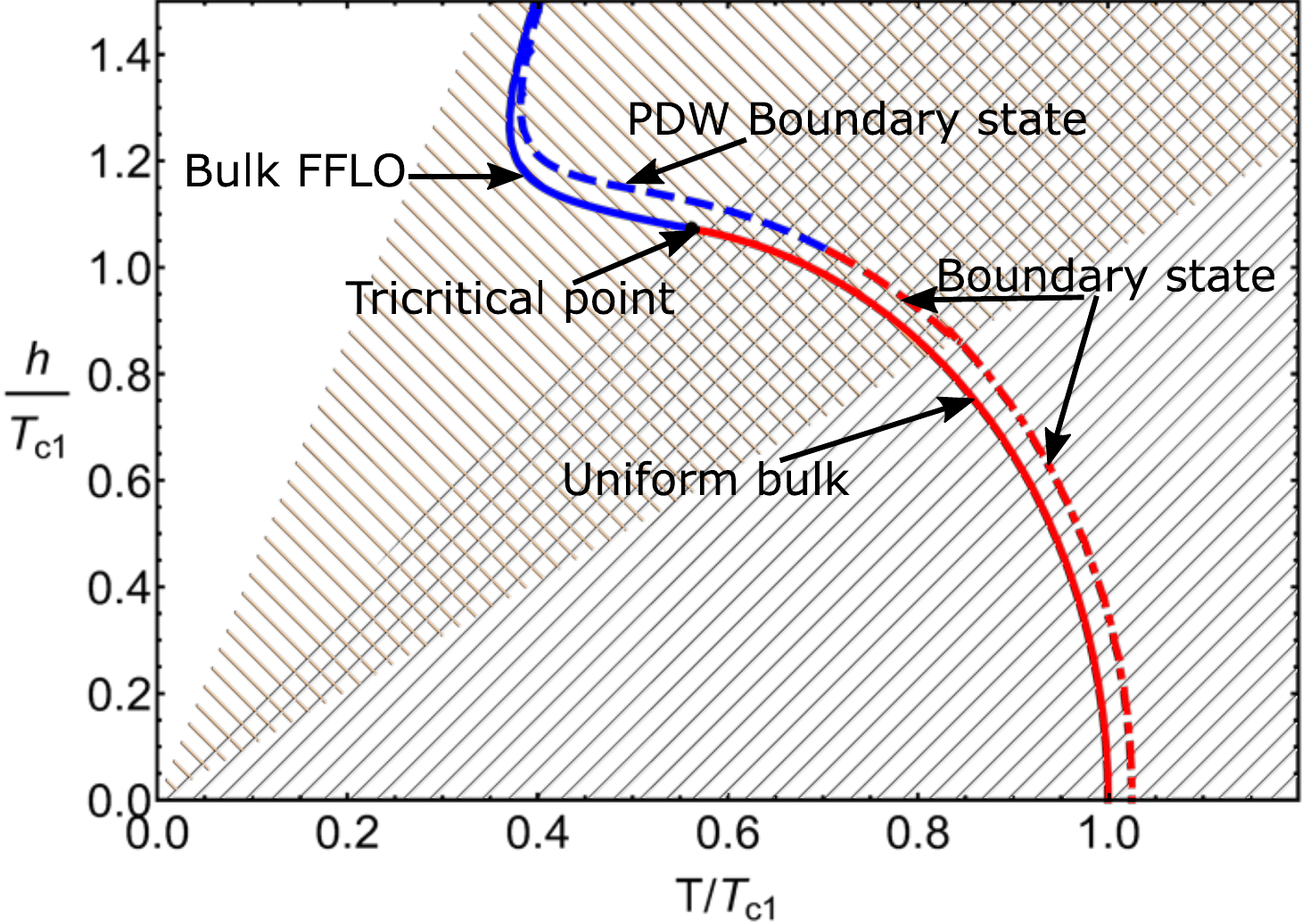}
\caption{
Phase diagram of a two-dimensional spin imbalanced superconductor in the GL model as a function of temperature $T$ and spin imbalance $h$ for $\gamma = - 0.025 \frac{N v_F}{T_{c1}}$ (for example, a system in an in-plane magnetic field).
Lines denote superconducting phase transitions.
The bulk phase transition between normal and superconducting states is denoted by solid lines according to \eqref{phase_transition_bulk}.
The boundary remains superconducting for higher temperatures.
The dashed lines denote a phase transition from surface superconducting to normal state according to \eqref{phase_transition_boundary_h} and the dot-dashed line according to \eqref{phase_transition_boundary}.
When the spin imbalance $h$ is large enough, bulk and boundary states turn from sign definite (red) to periodically modulated in space states (blue).
Gray hatching denotes the region where the usual GL model \eqref{F_GL} is bounded from below (it works best for $T \to T_{c1}$ and $h \to 0$).
Brown hashing shows the region where the GL model with higher-order derivatives \eqref{F_GL_h} is convergent (the best at the tricritical point).
Note, that this phase diagram has a boundary PDW state extending beyond the tricritical point and there is a smooth transition to a non-PDW boundary state, which agrees with the phase diagram obtained in a Bogoliubov-de Gennes formalism \cite{samoilenka2020pair}.
}
\label{fig_GL_phase_diagram}
\end{figure}

This gives us that bulk transitions to the normal state at, see \figref{fig_GL_phase_diagram}:
\begin{equation}\label{phase_transition_bulk}
\begin{gathered}
\text{FFLO: } K < 0 \text{ and } 4 \widetilde{K} \alpha = K^2 \\
\text{Uniform: } K > 0 \text{ and } \alpha = 0
\end{gathered}
\end{equation}

\begin{figure}
\centering
\includegraphics[width=0.99\linewidth]{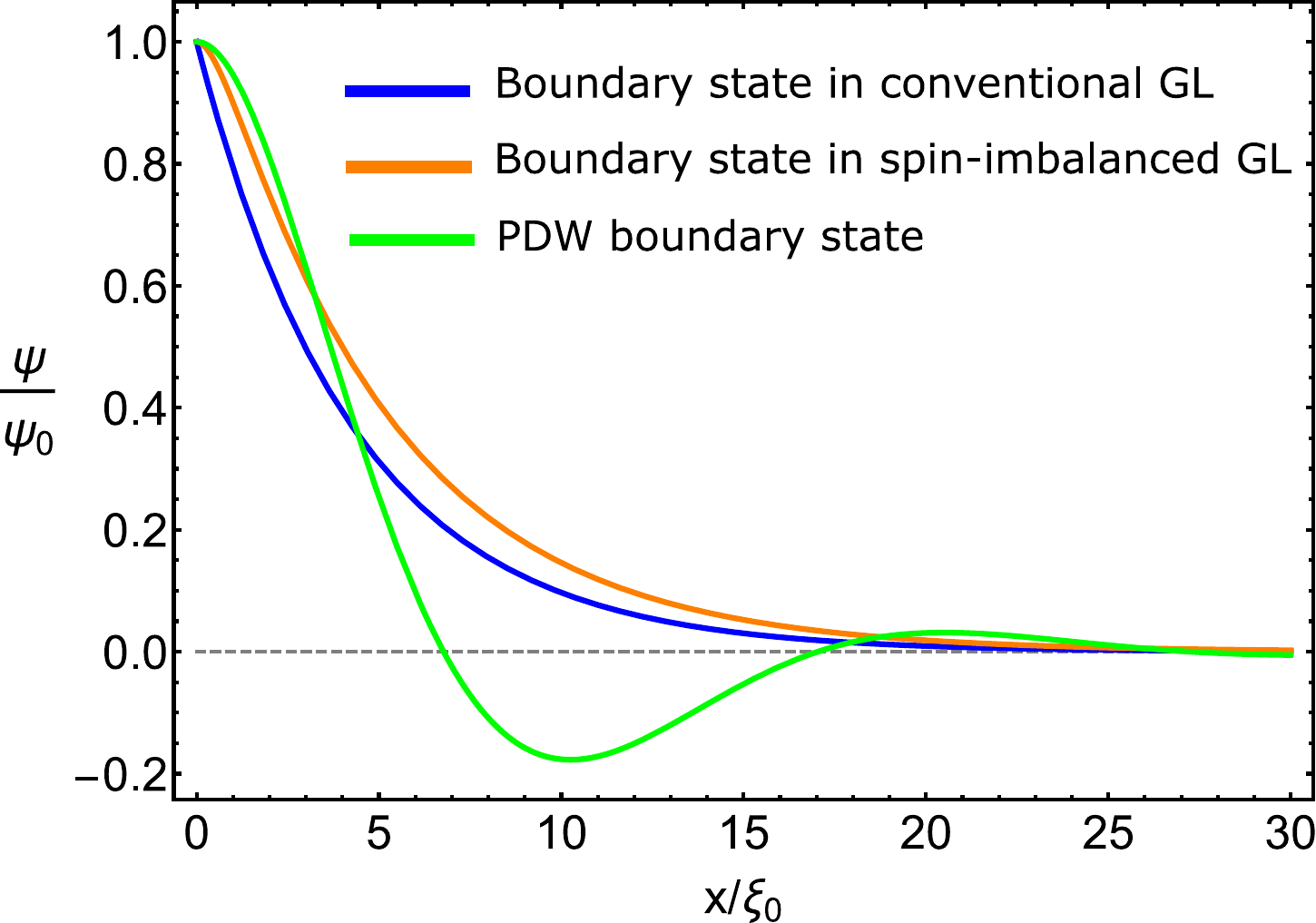}
\caption{
Solutions for the boundary state in the conventional GL model \eqref{F_GL} given by \eqref{boundary_psi} (blue) and in a spin-imbalanced GL model with higher-order derivatives \eqref{F_GL_h} given by \eqref{boundary_psi_h} (orange).
Both solutions are obtained for parameters $T / T_{c1} = 0.8$ and $h / T_{c1} \simeq 0.9$, such that they lie on the transition line in \figref{fig_GL_phase_diagram}.
The pair-density-wave (PDW) boundary-state solutions obtained in the GL model \eqref{F_GL_h} given by \eqref{boundary_psi_h} (green) at $T / T_{c1} = 0.5$ and $h / T_{c1} \simeq 1.1$.
The solutions are normalized to $\psi_0$, which is the $\psi$ value at the boundary and $\xi_0 = \frac{v_F \sqrt{7 \zeta(3)}}{4 \pi T_{c1} \sqrt{2}}$.
}
\label{fig_GL_psis}
\end{figure}

The system however has a boundary superconducting state with critical temperature which is higher than the bulk critical temperature.
Let us now consider the phase transition from a superconducting boundary state to a normal state.
For that matter consider sample positioned at $x > 0$.
Then \eqref{GL_eq_h_lin} should be solved together with the boundary conditions \eqref{bc_h} and the requirement that the order parameter goes to zero at infinity.
We obtain the order parameter configuration of the boundary state, see \figref{fig_GL_psis}:
\begin{equation}\label{boundary_psi_h}
\begin{gathered}
\psi = \text{const} \left( \frac{e^{ - q_+ x}}{q_+} - \frac{e^{ - q_- x}}{q_-} \right) \\
q_\pm = \sqrt{\frac{K \pm \sqrt{K^2 - 4 \widetilde{K} \alpha}}{2 \widetilde{K}}}
\end{gathered}
\end{equation}
which satisfies the second condition in \eqref{bc_h} at the transition from the superconducting boundary to the normal state, see \figref{fig_GL_phase_diagram}:
\begin{equation}\label{phase_transition_boundary_h}
\sqrt{\widetilde{K} \alpha} \left( q_+ + q_- \right) = - \gamma
\end{equation}

\section{Boundary states in a conventional superconductor}\label{sec_GL}
Next, we consider solutions for the conventional GL model \eqref{F_GL} with that derived in the above boundary conditions \eqref{bc_Lambda}.
The model can be obtained from the more general expression \eqref{F_GL_h} by setting $\widetilde{K} = K_1 = \nu = 0$.
Hence bulk transition from superconducting to normal state takes place at $\alpha = 0$, see \eqref{phase_transition_bulk}.
When the derived in the above microscopic boundary conditions are used the model exhibits superconducting boundary states.
The transition from it to a normal state is obtained from the boundary conditions \eqref{bc_Lambda} and \eqref{GL_eq_h_lin} at $\widetilde{K} = 0$.
The superconducting boundary-state solution that follows from that equation is:
\begin{equation}\label{boundary_psi}
\psi = \text{const} e^{ - q_0 x},\ \ q_0 = \sqrt{\frac{\alpha}{K}}
\end{equation}
Hence transition to a normal state is given by the following condition, which corresponds to the boundary condition \eqref{bc_Lambda}:
\begin{equation}\label{phase_transition_boundary}
\sqrt{\alpha K} = - \gamma
\end{equation}

This is the GL approximation of the superconducting boundary state obtained earlier as a solution of the full microscopic theory \cite{samoilenka2020boundary}.
Note, that the coarse-grained GL field is smoothly varying and the enhanced pairing correlations are modeled as a source in the form of a boundary integral, yielding the boundary conditions \eqref{bc_Lambda}.
Note, that this condition can be obtained from \eqref{phase_transition_boundary_h} by setting $\widetilde{K} \to 0$.
The latter also leads to $q_- \to q_0$ and $q_+ \to \sqrt{K / \widetilde{K}}$, which means that \eqref{boundary_psi_h} becomes \eqref{boundary_psi}.

From \eqref{phase_transition_boundary} we obtain that for zero population imbalance, the boundary state transitions to normal at $T_{c2}$ that is larger than bulk critical temperature $T_{c1}$:
\begin{equation}
\tau \equiv \frac{T_{c2} - T_{c1}}{T_{c1}} \simeq \ln \frac{T_{c2}}{T_{c1}} = \frac{\gamma^2}{N K}
\end{equation}

Hence, using \eqref{gamma_1d} for a one dimensional system, $\tau$ is equal:
\begin{equation}
\tau \simeq \frac{7 \zeta(3)}{4} \left( \frac{N V}{1 - \frac{N V}{2}} \right)^2
\end{equation}

For $N V \to 0$ this is equal to $\tau = \frac{7 \zeta(3)}{4} (N V)^2$, which we previously obtained in a one dimensional model without the Debye frequency \cite{samoilenka2020boundary} (note that, in \cite{samoilenka2020boundary} there is a typo and the rescaled interaction is actually $\hat{V} = N V$).

For two a dimensional system from \eqref{gamma_2d} we obtain:
\begin{equation}
\tau \simeq \frac{8}{7 \zeta(3)} \left( \frac{T}{k_F v_F} \frac{\ln \frac{k_F}{k_D}}{N V} \right)^2
\end{equation}

For a three-dimensional anisotropic system with $a > 1$, we obtain from \eqref{gamma_3d_a}:
\begin{equation}
\tau \simeq \frac{3}{7 \zeta(3)} \left( \frac{(a - 1) \pi^3 T}{k_F v_F N V} \right)^2
\end{equation}

\subsection{Analogy to wetting}
As a side note, one can imagine that the plot of $\psi$ in \figref{fig_GL_psis} for the usual GL model is a vertical cross-section of a tank filled with water, with $\psi$ being the height of the surface of the water.
This is not a coincidence.
The energy of a thin column of water of width $dx$ and height $\psi$ is composed of surface-tension energy $\sigma dl$ and gravitational energy $\frac{\rho g}{2} \psi^2 dx$,
where $\sigma$ is the energy per unit surface, $dl$ is the length of the surface, $\rho$ is the density of water, and $g$ is the gravitational constant. Note, that we implicitly redefined $\psi \to \lambda \psi$, where $\lambda$ is some dimensional constant, so that for new $[\psi] = [x]$. 
Then the total energy is
\begin{equation}
E \simeq \text{const} + \int_{0}^{L} dx \left( \frac{\sigma}{2} \left(\frac{d\psi}{dx}\right)^2 + \frac{\rho g}{2} \psi^2 \right)
\end{equation}

which is similar to the GL model \eqref{F_GL}, if we substitute $\sigma \to 2 K$ and $\rho g \to 2 \alpha$.

Then the problem of the boundary states in a superconductor can be related to the problem of adhesion of water to a wall.
In the latter case, the boundary condition is set by fixed contact angle, which is equivalent to fixing $ \textbf{n} \cdot \nabla \psi = \text{const} $.
This boundary condition is similar to \eqref{bc_Lambda} for a superconductor.
Note that in this analogy, the superconductor-insulator interface behaves like a hydrophilic surface.
For other interfaces, like superconductor-normal metal, or an interface with certain types of different boundary layers \cite{samoilenka2020boundary}, the gap can be suppressed near the boundary, which corresponds to a hydrophobic surface.

\section{$H_{c3}$ from microscopically derived GL theory revisited}\label{sec_GL_H}
For conventional boundary conditions $\gamma=0$, superconductivity in type-II materials survives near surfaces at the magnetic fields up to $H_{c3}$, which is higher than the critical field associated with the disappearance of superconductivity in the bulk $H_{c2}$.
The boundary conditions that we derived have direct implications for the third critical magnetic field $H_{c3}$.

In the conventional picture \cite{saint1963onset} it is described by solving the linearized GL equation by using the standard de Gennes boundary conditions $\textbf{n} \cdot \nabla \psi = 0$.
Note that it has been observed earlier that numerical solution in the fully microscopic Bogoliubov-de Gennes theory (i.e., obtained beyond quasiclassical approach) is not consistent with this picture \cite{troy1995self}, but should be more robust, however, that work did not determine $H_{c3}$.
We are in a position now to calculate $H_{c3}$.
Note that the problem that we will study below, namely $H_{c3}$ in a GL model with an included surface term have been studied on phenomenological grounds in the past in the context of superconductors with modified surface layers, and superconductors with enhanced superconductivity on twinning planes \cite{fink1969surface,buzdin1981localized,averin1983theory,khlyustikov1987twinning}.
Remarkably, enhanced $H_{c3}$ was observed for ordinary surfaces of the elemental superconductors and experimental papers have been explicitly raising the question of whether that originates in enhanced superconducting properties of surfaces of unknown origin \cite{khlyustikov2011critical,khlyustikov2016surface}.
Our goal here is to calculate $H_{c3}$ in the microscopically derived GL theory corresponding to a regular boundary of a standard BCS superconductor.

To include magnetic field one simply replaces the derivative with a covariant derivative in \eqref{FermiHubbard} as $\nabla \to \nabla - \ii e \textbf{A}$, where $\textbf{A}$ is the magnetic vector potential and $e$ is the electron charge.
Next, Green's function with magnetic field $G_\sigma^A$ is obtained as
\begin{equation}
G_\sigma^A(\textbf{r}, \textbf{r}') = e^{\ii \phi(\textbf{r}, \textbf{r}')} G_\sigma(\textbf{r}, \textbf{r}')
\end{equation}

where $G_\sigma$ is the Green's function for zero magnetic field defined in \eqref{G_def} and $\phi(\textbf{r}, \textbf{r}') \simeq e \textbf{A} \cdot (\textbf{r} - \textbf{r}')$.
Note that, in this approximation, it is assumed that $\textbf{A}$ is very slowly changing and hence it can be $\textbf{A} \simeq \textbf{A}(\textbf{r})$ or $\textbf{A} \simeq \textbf{A}(\textbf{r}')$ or anywhere close to $\textbf{r},\ \textbf{r}'$.
For details see \cite{gor1960critical,abrikosov1965concerning,Abrikosov_Gorkov_Dzyaloshinski_Book}.
In our calculation, however, it is convenient to choose
\begin{equation}
\phi(\textbf{r}, \textbf{r}') \simeq e \textbf{A}(\textbf{r}) \cdot \textbf{r} - e \textbf{A}(\textbf{r}') \cdot \textbf{r}'
\end{equation}
and to extend $\textbf{A}(\underline{\textbf{r}}) = \underline{\textbf{A}}(\textbf{r})$\footnote{Note, that this extension is purely fictitious and is used just to derive the boundary conditions for $\Delta$ easily. The actual $\textbf{A}$ has different value outside of the superconductor, which follows from the magnetic energy defined over all space, $\int_{- \infty}^{+\infty} d\textbf{r} \frac{(\textbf{B} - \textbf{H})^2}{2}$, where $\textbf{B} \equiv \nabla \times \textbf{A}$ and $\textbf{H}$ is the external magnetic field. }.

Then property \eqref{G_reflection} is satisfied for $G_\sigma^A$ as well.
It means that the derivation for the model with the magnetic field is similar to the one outlined in \secref{sec_derivation}.
The only difference is that now $\Delta$ is replaced by:
\begin{equation}
\Delta(\textbf{r}) \to \Delta^A(\textbf{r}) = \Delta(\textbf{r}) e^{- 2 \ii e \textbf{A}(\textbf{r}) \cdot \textbf{r}}
\end{equation}

Hence, for the resulting GL model the magnetic field amounts to replacing
\begin{equation}
\nabla \to \nabla - 2 \ii e \textbf{A}
\end{equation}

In an external magnetic field, the boundary condition with $\gamma < 0$ leads to an increased critical magnetic field $H_{c3}$.
It has a different dependence on temperature compared with the standard textbook derivation \cite{saint1963onset}, see,for example, the phenomenological discussion in \cite{fink1969surface}.
Similar observations were made in phenomenological studies of superconductors with twinning planes.

\begin{figure}
\centering
\includegraphics[width=0.99\linewidth]{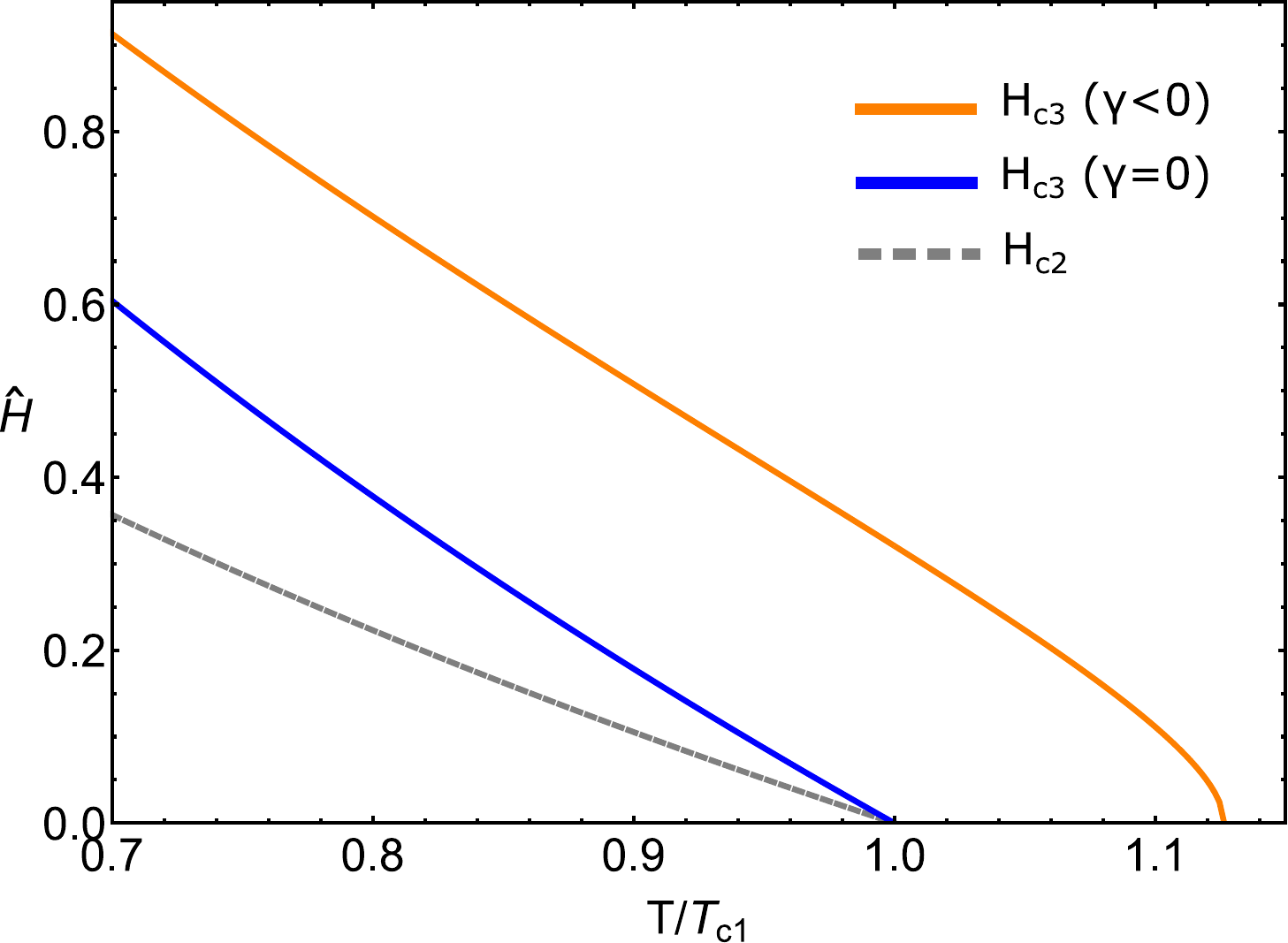}
\caption{
Rescaled critical magnetic fields $\hat{H} \equiv \frac{2 |e| K}{N} H$ as a function of temperature $T$.
Bulk transitions to the normal state for fields higher than $H_{c2} = - \frac{\alpha}{2 |e| K}$.
In the derivation \cite{saint1963onset} with zero boundary term ($\gamma = 0$), the surface transitions to the normal state at higher field $H_{c3} \simeq 1.69 H_{c2}$.
From \eqref{Hc3_sol} we obtain that, if in \eqref{F_GL} the boundary term is present with $\gamma < 0$, then surface superconductivity is enhanced and boundary transitions to normal at field $H_{c3}$ higher than that for $\gamma = 0$.
Here we have chosen $\gamma = - 0.05 \frac{N v_F}{T_{c1}}$.
}
\label{fig_GL_Hc3}
\end{figure}

Here we compute $H_{c3}$ with the microscopic boundary conditions derived above.
Consider a two- or three-dimensional system and assume that the external magnetic field is directed along the $z$ direction and equals $H$.
Hence we can set a gauge for vector potential so that only nonzero component is $A_y$.
Then the transition to the normal state is obtained by solving the linearized GL equation in terms of $\psi(x)$:
\begin{equation}\label{GL_eq_A}
\alpha \psi - K \partial_x^2 \psi + K (2 e A_y)^2 \psi = 0
\end{equation}
together with the boundary conditions \eqref{bc_Lambda}.
We set $A_y = (x + x_0) H$, where $x_0$ is to be optimized to get the highest $H$.
Then the solution to \eqref{GL_eq_A} is
\begin{equation}\label{psi_sol_Hc3}
\psi = \text{D}_{- \frac{1}{2} \left(1 + \frac{\alpha}{2 |e| H K} \right) }\left( (x + x_0) \sqrt{4 |e| H} \right)
\end{equation}
where $\text{D}_\nu(x)$ is parabolic cylinder function.
To find $x_0$, it is convenient to calculate the first $d\psi$ integral of \eqref{GL_eq_A}.
By using boundary conditions \eqref{bc_Lambda}, we obtain
\begin{equation}\label{int_eq_Hc3}
\psi^2(x_0) \left[ \frac{\gamma^2 - \alpha K}{K^2} - (2 e H x_0)^2 \right] - \int_{x_0}^{\infty} (2 e H \psi(x))^2 2 x dx = 0.
\end{equation}
This equation sets the relation between $x_0$ and $H$, when the solution for $\psi$ \eqref{psi_sol_Hc3} is inserted.
Since we search for the largest $H$, a derivative of \eqref{int_eq_Hc3} with respect to $x_0$ should be zero.
It gives:
\begin{equation}\label{x_0_sol}
x_0 = - \frac{\sqrt{\gamma^2 - \alpha K}}{2 |e| H K}.
\end{equation}
Here we picked the minus sign so that \eqref{int_eq_Hc3} can be satisfied.
Next, in order to find $H$ we can either solve \eqref{int_eq_Hc3} or the boundary condition \eqref{bc_Lambda}.
We solve the latter, which using the expression for $x_0$ \eqref{x_0_sol} and rescaling, amounts to solving numerically for $\eta$ for a given $a$ in:
\begin{equation}\label{Hc3_sol}
\frac{(1 - \sqrt{a})}{\sqrt{\eta}} \text{H}_{\frac{a - 1 - \eta}{2 \eta}}\left( - \sqrt{a / \eta} \right) = \text{H}_{\frac{a - 1 + \eta}{2 \eta}}\left( - \sqrt{a / \eta} \right)
\end{equation}
where $\text{H}_\nu(x)$ is a Hermite polynomial, $a = 1 - \frac{\alpha K}{\gamma^2}$ and $\eta = \frac{2 |e| H K^2}{\gamma^2}$.
The $H$ obtained then is equal to $H_{c3}$.
See \figref{fig_GL_Hc3} for a plot of $H_{c3}$.
The $H_{c3}$ is, therefore, higher than in the original Saint-James de Gennes derivation \cite{saint1963onset}.
Also, note that $H_{c3}$ should exist for type-I superconductors in significant temperature range.

\section{Conclusions}
We considered the generic BCS model for spin-balanced and spin-imbalanced fermions.
From that model, we derived the boundary conditions for the GL theory for the interface between a superconductor and an insulator.
We showed that the free energy of a superconductor acquires an additional term given by the surface integral of $\gamma |\psi|^2$.
The physical origin of this term is the fact that near a well-reflecting boundary, the total gap oscillates with $\simeq k_F$ momentum.
This oscillatory part is coupled to the averaged gap $\psi$ in GL theory, leading to an additional surface term.
We obtained that $\gamma < 0$ for one- and two-dimensional continuous BCS models.
For the three-dimensional isotropic BCS model, the surface term is beyond the resolution of our analytical approach.
Whereas for an anisotropic three-dimensional model we have shown that $\gamma$ can be positive or negative. 
The negative $\gamma$ leads to enhanced superconductivity near boundaries.
Note that, for the tight-binding BCS model the surface superconductivity exists in all dimensions \cite{samoilenka2020boundary}, which also implies boundary conditions with $\gamma<0$.

To obtain boundary conditions we showed that the following procedure can be applied.
Consider the GL model that in general has the highest in derivatives term of order $k$.
Then one should reflect the order parameter in the boundary and write the free energy as an integral over the whole space.
This reflection automatically applies proper boundary conditions.
Namely, one can write kinetic terms in any integrated-by-parts form and search for a minimum of the total free energy.
Boundary conditions are obtained from the condition on energy to be finite and from the variation of this functional with respect to the order parameter. 
As a result, we obtained that usually all normal to boundary odd derivatives of $\psi$ of the order less than $k - 1$ should be zero, whereas $k - 1$ derivative will be proportional to $\gamma \psi$.

For the standard GL model with second-order derivatives, $k = 2$ and hence we obtain boundary conditions $\textbf{n} \cdot \nabla \psi = - \frac{\gamma}{K} \psi$, see \eqref{bc_Lambda}.
GL model for spin-imbalanced systems requires taking into account fourth-order gradient terms and hence has boundary conditions $\textbf{n} \cdot \nabla \psi = 0,\ \textbf{n} \cdot \nabla^3 \psi = \frac{\gamma}{\widetilde{K}} \psi$, see \eqref{bc_h}.
Using these new microscopically derived boundary conditions we obtained superconducting boundary states in the conventional GL model and revised the calculation of PDW boundary states in the spin-imbalanced GL model.

The obtained boundary conditions allow GL theory to account, in a microscopically accurate way for boundary states that were found earlier in fully microscopic solutions of BCS theory \cite{samoilenka2020boundary}.
Namely in the GL model for $\gamma < 0$, in zero external magnetic field, the superconducting gap is larger near the surface than in the bulk, and superconductivity survives for higher temperatures.
Since microscopic calculations show that superconductivity is more enhanced at the edges and corners of a three-dimensional sample, to model these effects one should add analogous extra contributions for corners and edges.

By adding an external magnetic field we revised the theory of the third critical magnetic field $H_{c3}$ for a BCS superconductor.
The surface effects make this field larger and extending in a type-I regime, compared with results obtained using de Gennes boundary conditions. 

We note, that these surface effects can be described in a quasiclassical approach if one augments the theory by taking into account higher-momentum contributions to the boundary conditions.

\begin{acknowledgments}
We thank Andrea Benfenati and Mats Barkman for the discussions.
The work was supported by the Swedish Research Council Grants No. 642-2013-7837, 2016-06122, 2018-03659, the G\"{o}ran Gustafsson Foundation for Research in Natural Sciences and Medicine.
\end{acknowledgments}


%

\end{document}